# High-pressure study of substrate material ScAlMgO$_4$

D. Errandonea[1], R. S. Kumar[2], J. Ruiz-Fuertes[1], A. Segura[1], E. Haussühl[3]

[1]MALTA Consolider Team, Departamento de Física Aplicada - ICMUV, Universitat de València, Edificio de Investigación, c/Dr. Moliner 50, 46100 Burjassot, Valencia, Spain

[2]High Pressure Science and Engineering Center, Department of Physics and Astronomy, University of Nevada Las Vegas, 4505 Maryland Parkway, Las Vegas, Nevada 89154-4002, USA

[3]Institut für Geowissenschaften, Abt. Kristallographie, Goethe-Universität Frankfurt, Altenhöferallee 1, D-60438 Frankfurt am Main, Germany

**Abstract**

We report on the structural properties of ScAlMgO$_4$ studied under quasi-hydrostatic pressure using synchrotron high-pressure x-ray diffraction up to 40 GPa. We also report on single-crystal studies of ScAlMgO$_4$ performed at 300 K and 100 K. We found that the low-pressure phase remains stable up to 24 GPa. At 28 GPa, we detected a reversible phase transformation. The high-pressure phase is assigned to a monoclinic distortion of the low-pressure phase. No additional phase transition is observed up to 40 GPa. In addition, the equation of state, compressibility tensor, and thermal expansion coefficients of ScAlMgO$_4$ are determined. The bulk modulus of ScAlMgO$_4$ is found to be 143(8) GPa, with a strong compressibility anisotropy. For the trigonal low-pressure phase, the compressibility along the $c$-axis is twice than perpendicular one. A perfect lattice match with ZnO is retained under pressure in the pressure range of stability of wurtzite ZnO.

## I. Introduction

In recent years GaN and ZnO have attracted increasing attention due to their potential applications in optoelectronic devices [1]. The progress in the developing of these devices is constrained by availability of suitable substrate materials. The main factors determining the appropriate substrate material are matched lattice parameters, thermal expansion and compressibility. Scandium magnesium aluminate ($ScAlMgO_4$), due its excellent lattice matching with GaN and ZnO (lattice mismatch 1.8% and 0.09%), is one of the most promising substrate material for these semiconductors [2]. $ScAlMgO_4$ was first synthesized by Kimuzuka and Mohri [3], who determined that it has a rhombohedral structure ($R\bar{3}m$) similar to that of $YbFeO_4$; see Fig. 1. More recently, a single-crystal x-ray diffraction study was performed on $ScAlMgO_4$, but the atomic positions remain unpublished [4]. In addition, no information on the axial and bulk compressibility of $ScAlMgO_4$ has been reported yet. Very thin substrates can be prepared owing to the easy cleaving of crystalline $ScAlMgO_4$ along planes perpendicular to the *c*-axis, which constitutes a further advantage for high-pressure optical studies on GaN or ZnO thin films and quantum structures. It is then interesting to investigate the evolution of its lattice matching under pressure. Here we characterize the ambient and high-pressure crystal structure of $ScAlMgO_4$ by using high-resolution powder and single-crystal x-ray diffraction techniques. We also study the effects of pressure on it. We found that compression is anisotropic and determined the room-temperature (RT) equation of state (EOS). We also discovered a new pressure-induced structural phase transition at 28 GPa.

## II. Experimental details

The studies were performed on samples obtained from single-crystal substrates provided by MTI corporation. The structure $ScAlMgO_4$ was determined at ambient conditions by single-crystal diffraction. Single-crystal x-ray diffraction data were collected at 100 K and at room temperature, using an Xcalibur3 4-circle diffractometer from Oxford diffraction with a charge-coupled device (CCD) camera and Mo $K_\alpha$ radiation from a Mo anode operating at 45 kV and 38 mA. The diffractometer is equipped with a cryostream system (cryojet HT Oxford diffraction). This allows us to maintain the sample at a minimum temperature of 100 K within an error of 2 K throughout the measurement. We used a single crystal of 100x87x60 $\mu m^3$ mounted at a distance of 4.2 cm from the detector. We collected 837 frames with a frame width of 1° and exposure time of 60 s. Data reduction and absorption corrections using rhombohedral Laue symmetry for corrections were performed using the program CRYSALIS. As the starting atomic positions that we used were those obtained by the Rietveld refinement of the powder x-ray pattern measured by us at ambient conditions. The structure refinements were carried out with SHELXL97-2.

In order to determine if $ScAlMgO_4$ has a center-symmetric structure, as reported in Ref. [3] we applied the second-harmonic generation (SHG) technique employing an infrared IR laser (Falcon 217D, Quantronix), operating at $\lambda = 1054$ nm with a repetition of 1 KHz and a pulse width of 130 ns. The intensity of the 1 W laser power was decreased with an absorption filter (optical density 0.7). For the detection of a possible SHG signal a photomultiplier (R2949, Hamamatsu) and a photon counter (SRS4000, Stanford Research System) were used [5]. The sample was checked both with single crystal form and powder without obtaining any SHG signal.

Powder-diffraction studies were performed at ambient conditions in micron-size powder samples cleaved and ground from the single crystal. The measurements were carried out with a Seifert XRD 3003 TT diffractometer using Cu $K_\alpha$ monochromatic radiation ($\lambda = 1.5406$ Å). In order to perform high-pressure studies, pre-pressed pellets of ScAlMgO$_4$ were prepared using the finely ground powder obtained from the single crystal. Two independent experimental runs were performed up to 24 and 40 GPa. The powder samples were loaded in a 130 μm hole of a rhenium gasket pre-indented to 40 μm in a symmetric diamond-anvil cell (DAC) with diamond-culet sizes of 350 μm. A few ruby grains were loaded with the sample for pressure determination [6] and neon (Ne) was used as pressure-transmitting medium [7, 8]. At pressures higher than 4 GPa the EOS of Ne was used to double-check the pressure [9]. Pressure differences between both methods were always smaller that 0.2 GPa. Angle-dispersive x-ray diffraction (ADXRD) experiments were carried out at Sectors 16-BMD and 16-IDB of the HPCAT, at the Advanced Photon Source (APS), with an incident wavelength of 0.41514 Å in one experiment and of 0.40753 Å in the other. The monochromatic x-ray beam was focused down to 10x10 μm$^2$ using Kickpatrick–Baez mirrors. The images were collected using a MAR345 image plate located 383 mm (or 350 mm) away from the sample and then integrated and corrected for distortions using FIT2D [10]. The structure solution and refinements were performed using the POWDERCELL [11] and GSAS [12] program packages.

III.    **Results and discussion**

A.    **Ambient pressure structure**

ScAlMgO$_4$ has been reported to have a rhombohedral center-symmetric structure ($R\bar{3}m$) [3, 4, 13]. At the detection limit of our setup we did not observe any SHG, which in principle is consistent with the center-symmetric character of the crystal structure. Powder and single-crystal diffraction confirmed the assignation of the space group $R\bar{3}m$. After a Rietveld refinement of a powder x-ray diffraction pattern collected at ambient pressure (0.1 MPa) outside the DAC the following structural parameters for ScAlMgO$_4$ were obtained: $a$ = 3.245(1) Å and $c$ = 25.160(9) Å. The structure has three formula units per unit cell (Z = 3) and the unit-cell volume is 229.4(2) Å$^3$. The refinement residuals are $R_F^2$ = 2.26%, $R_{WP}$ = 3.57%, and $R_P$ = 1.86%. The atomic positions, obtained for the structure, are summarized in Table I. Single-crystal diffraction provides similar atomic positions and unit-cell parameters (see Table II). In this case a total of 1178 reflections were measured (103 unique reflections). More details of data collection and agreement factors are given in Table II. The obtained unit-cell parameters agree among themselves and better with those reported by Tang *et al.* [13] (*a* = 3.2459 Å, *c* = 25.1602 Å) than with those reported by Zhou *et al.* [4] (*a* = 3.2506 Å, *c* = 25.152 Å). The ambient pressure structure is illustrated in Fig. 1. It is built by stacking of oxygen atoms along the *c*-axis with a closest packing topology. Sc is located between two oxygen plans in octahedral coordination, whereas Al/Mg is almost in the same plane as oxygen atoms, in trigonal bipyramidal coordination. Basically the structure can be described as [AlMgO$_4$]$^{3-}$ layers parallel to the *ab* plane connected into a three-dimensional framework by the Sc atoms via an oxygen atom. The Sc-O distance is 2.133(1) and the Al/Mg-O distances are 1.898(1) Å, 1.905(1) Å, and 2.236(1) Å (see Fig. 1). The second bond, 1.905(1) Å, is between Al/Mg and the oxygen connecting the [AlMgO$_4$]$^{3-}$ layers with Sc, and it is oriented along the *c*-axis. The other two distances correspond to bonds within the [AlMgO$_4$]$^{3-}$ layers. Three short bonds in the *ab* plane and a longer bond perpendicular to it.

Single-crystal diffraction at 100 K shows that the crystal structure of ScAlMgO$_4$ is the same that at 300 K. In this case a total of 1339 reflections were measured (111 unique reflections). More details of data collection and agreement factors are given in Table III. The obtained unit-cell parameters indicate that the thermal expansion is slightly larger along the *c*-axis than along *a*-axis. The linear thermal expansion coefficients are 8.88(6) 10$^{-6}$ K$^{-1}$ and 7.68(5) 10$^{-6}$ K$^{-1}$, respectively. The relative volume reduction from 300 K to 100 K is 0.4%. In addition, no important changes are induced in the atomic positions upon cooling (see Tables II and III).

## B. High-pressure studies of the low-pressure phase

A summary of the results obtained in one of the high-pressure x-ray diffraction experiments performed for ScAlMgO$_4$ up to 24 GPa is shown in Fig. 2. We did not find any evidence of the occurrence of structural changes. The results can be summarized as follows. At 4.5 GPa we observed the appearance of diffraction peaks due to the solidification of Ne [9]. These peaks can be identified since Ne is much more compressible than ScAlMgO$_4$, and therefore the Ne peaks have a different pressure evolution that the Bragg peaks of the sample (see Fig. 2). In addition, not all the peaks of ScAlMgO$_4$ move in the same way under compression. The (00$l$) reflections; e.g. (006) and (009), move towards higher angles with higher pressure rate than the rest of the reflections. This can be seen in Fig. 2 by comparing the pressure evolution of (006) and (104) Bragg peaks. This fact is indicating a differential axial compressibility in ScAlMgO$_4$. This phenomena is also illustrated by the two reflections located around 2θ = 8.5º. At ambient pressure there is a strong peak corresponding to (101) and (009) reflections and on the right hand side of it a weaker peak associated to the (012) reflection. As pressure increase, the (009) peak moves considerably more than the others. As a consequence of it, first the strong peak splits into two peaks (see the spectrum collected at 7.5 GPa) and consequently the (009) peak merges with the (012) reflection. This causes a gradual change of the intensity of the peaks located around 2θ = 8.5º as shown in Fig. 2. Another evidence of the differential axial compressibility of ScAlMgO$_4$ is the splitting under compression of (018) and (0012) reflections and (116) and (0114) reflections. It is important to comment here that the width of Bragg peaks does not considerable change under compression and all peaks are well resolved up to 24 GPa. This fact suggests that the use of neon as pressure medium creates quasi-hydrostatic conditions in the whole pressure range [7] avoiding therefore any influence of uniaxial stresses on the reported results [14].

From the Rietveld refinement of x-ray diffraction patterns we have obtained the pressure dependence of the lattice parameters of ScAlMgO$_4$. The pressure evolution of the structural parameters and the atomic volume (V) are shown in Fig. 3. There it can be seen that the *c*-axis is more compressible that the *a*-axis and that both axes have a non-linear pressure dependence, which becomes more evident beyond 10 GPa. As a consequence of the differential axial compressibility, the axial ratio decreases from 7.75 at ambient pressure to 7.56 at 24 GPa (see Fig. 3). At low pressure, the mean linear compressibilities of ScAlMgO$_4$ are $\beta_a$ = 1.56 (3) 10$^{-3}$ GPa$^{-1}$ and $\beta_c$ = 3.18(4) 10$^{-3}$ GPa$^{-1}$. The first value indicates that the compressibility in the [AlMgO$_4$]$^{3-}$ layers (*ab* plane) is similar to that of related covalent oxides like perovskite ScAlO$_3$ [15], spinel MgAl$_2$O$_4$ [16], and zircon-type ScVO$_4$ and ScPO$_4$ [8, 17]. On the contrary the compressibility along the *c*-axis is considerably larger indicating probably a weak bonding between the layers that constitute ScAlMgO$_4$.

It is important for high-pressure studies on GaN and ZnO to compare their mechanical properties with those of substrate ScMgAlO$_4$. In the pressure range up to 10 GPa (range of stability of wurtzite ZnO) the mean linear compressibility of ScAlMgO$_4$ is $\beta_a$ = 1.56(3) 10$^{-3}$ GPa$^{-1}$. This value is similar to the values of $\beta_a$ = 1.43(3) 10$^{-3}$ GPa$^{-1}$ and $\beta_a$ = 1.60(3) 10$^{-3}$ GPa$^{-1}$ of ZnO as obtained from x-ray diffraction and extended x-ray absorption fine structure (EXAFS) measurements under pressure respectively [18]. Then, in the pressure range of stability of wurtzite ZnO, its *a*-axis compressibility is virtually identical to the one of the *a*-axis of ScAlMgO$_4$, indicating that the lattice match does not practically change under pressure. Consequently ZnO thin films grown on ScAlMgO$_4$ can be compressed without being subjected to significant biaxial stress. In the case GaN [19, 20], with a smaller compressibility, the lattice mismatch with ScAlMgO$_4$ slightly decreases under pressure from 1.78% at ambient pressure to 1.45% at 10 GPa. This situation contrasts with the case of thin films of ZnO or GaN deposited on c-oriented sapphire for which the lattice mismatch increases under pressure due to the much smaller compressibility of sapphire.

In order to determine the EOS of ScAlMgO$_4$, the pressure-volume curves shown in Fig. 3 were analyzed using a third-order Birch–Murnaghan equation. The following parameters were obtained: $V_0$ = 229.3(7) Å$^3$, $B_0$ = 137(9) GPa, and $B_0$' = 8.3(9), being $V_0$, $B_0$, and $B_0$' the zero-pressure volume, bulk modulus, and pressure derivative of the bulk modulus, respectively. The obtained EOS is plotted in Fig. 3 together with the experimental data. The value determined for $B_0$' is larger than usual values in most substances (3.5 – 6.5) [21, 22]. This fact may reflect a gradual change in the compression mechanism over the pressure range we studied. In particular, the fact that the data point collected at 24 GPa deviates from the EOS fit supports this hypothesis. Consequently, we constrained the P-V data to a pressure range up to 18.5 GPa and obtained the following EOS parameters: $V_0$ = 229.4(6) Å$^3$, $B_0$ = 143(8) GPa, and $B_0$' = 5.9(7). This bulk modulus (143 GPa) is 25% smaller than that of spinel MgAl$_2$O$_4$, $B_0$ = 190 GPa [17], and 35% smaller than that of perovskite ScAlO$_3$, $B_0$ = 218 GPa [16]. In contrast ScAlMgO$_4$ has a bulk modulus similar to Sc$_2$O$_3$, $B_0$ = 154 GPa. The reason behind the smaller bulk modulus of ScAlMgO$_4$ compared with related oxides may be due to the large compressibility of the *c*-axis. This argument is consistent with the fact that Sc$_2$O$_3$, another compound with a layered structure, also behaves in a similar way.

In order to understand the non-isotropic compression of ScAlMgO$_4$, we extracted the pressure evolution of the atomic bonds from the structural refinements.

The results are shown in Fig. 4. There it can be seen that one Al/Mg-O distance is considerable less compressible than the other bond distances. This distance (the shortest one at ambient pressure) corresponds to bonds within the *ab* plane. On the other hand, the other two Al/Mg-O bonds are the most compressible bonds. In particular, the interlayer Al/Mg-O bond becomes the shortest one beyond 7 GPa (see Fig. 4). In contrast, the Sc-O bonds are slightly less compressible. We would like to note here that the $ScO_6$ octahedra do not distort upon compression. From this picture, we can conclude that the reduction of the two vertical Al/Mg-O bonds is what makes the *c*-axis the most compressible one. In contrast, the planes perpendicular to this direction are highly incompressible because the short Al/Mg-O bonds aligned along these planes are quite strong. This could be probably related to a preferred directionality of the valence-electron density as happens in layered $Ni_2Si$ [23].

### C. Phase transition

We will discuss now structural changes found beyond 24 GPa. Figure 5 compares two diffraction patterns measured at 24 and 28 GPa. We found that at 28 GPa important changes take places in the diffraction pattern. In particular, the relative intensity of the two strong peaks located near $2\theta = 8.5°$ changes, the least intense peak at 24 GPa becomes the most intense at 28 GPa. Also, there is an extra peak clearly emerging near $2\theta = 10°$. In addition, many other peaks arise and the diffraction peaks broaden. For the low-pressure phase we identified 48 reflections while for the high-pressure one we identified up to 97. In contrast with the sample peaks, the Ne reflections only move towards high angles as a consequence of the pressure increase. All the changes observed in the diffraction patterns indicate a pressure-induced phase transition occurring at 28 GPa. Upon further compression up to 40 GPa, there are no additional changes in the diffraction pattern with the exception of the peaks displacement due to the unit-cell parameters reduction (see Fig. 5). We also found that upon decompression the observed changes are reverted, indicating that the structural phase-transition is reversible. In particular, in Fig. 5 it can be seen that the diffraction pattern collected at ambient pressure upon decompression is very similar to the one shown in Fig. 2 corresponding to ambient pressure.

In an attempt to identify the structure of the high-pressure phase we considered several subgroups of space group $R\bar{3}m$. We found that a monoclinic structure with space group *C2/m* can satisfactory explain the diffraction patterns measured at 28 GPa and higher pressures. This structure can be obtained through a translationsgleiche transformation from the low-pressure structure and reduced to it for a given selection of structural parameters. From the diffraction pattern measured at 28 GPa we obtained for the high-pressure phase the following information: space group *C2/m*, Z = 2, $a$ = 16.07 Å, $b$ = 3.15 Å, $c$ = 8.02 Å, $\beta$ = 160.8°, V = 133.52 Å$^3$. The atomic positions for this structure are given in Table IV. At 40 GPa we obtained, for the same structure, the following structure parameters $a$ = 15.75 Å, $b$ = 3.09 Å, $c$ = 7.86 Å, $\beta$ = 161°, V = 124.50 Å$^3$. Apparently there is no volume discontinuity at the transition, which is consistent with the fact that the high-pressure phase can be obtained as a continuous distortion from the low-pressure structure. The proposed structure for the high-pressure phase is consistent with the behavior shown by other layered materials under compression, where pressure gradually changes the symmetry of the materials [24]. The fact that there is no abrupt change of the structure at the phase transition suggests that ScAlMgO$_4$ could be a good substrate to perform high-pressure experiments in ZnO and GaN even at pressures higher than the transition pressure. High-pressure single-crystal x-ray diffraction studies should be performed to confirm the proposed high-pressure structure of ScAlMgO$_4$. Regarding the compressibility of the high-pressure phase, we collected data for this phase only at four different pressures. Therefore, there is not enough information to accurately determine the axial and bulk compressibility of the high-pressure phase. However, we observed that the volume of the high-pressure phase can be reasonably well fitted by the EOS of the low-pressure phase, which suggests that both phases have similar bulk compressibility.

IV.  **Concluding Remarks**

We reported single-crystal x-ray diffraction and high-pressure powder diffraction studies of $ScAlMgO_4$ up to 40 GPa. We found that the low-pressure phase of $ScAlMgO_4$ reversibly transforms to another structure at 28 GPa. For the high-pressure phase we propose a monoclinic structure, which is a distortion of the low-pressure one. No additional transition is found up to 40 GPa. In addition, the EOS, compressibility tensor, and thermal expansion coefficients of $ScAlMgO_4$ are determined. The bulk modulus of $ScAlMgO_4$ is 143(8) GPa, with a strong compressibility anisotropy. Finally, the lattice mismatch of $ScAlMgO_4$ with semiconductors like ZnO and GaN is minimum in the pressure-stability range of the low-pressure phase. Therefore, $ScAlMgO_4$ constitutes and excellent substrate material to perform high-pressure optical studies on GaN or ZnO thin films and quantum structures up to 24 GPa.

**Acknowledgments**

We thank Dr. L. Bayarjargal for assistance in the SHG measurements. The authors thank for the financial support from Spanish MICCIN (Grants MAT2010-21270-C04-01, MAT2008-06873-C02-02 and CSD2007-00045) and the Deutsche Forschungsgemeinschaft (DFG) under project number HA 5137/3. J.R.F. is indebted to Spanish MEC for its support through an FPI fellowship. Portions of this work were performed at HPCAT (Sector 16), Advanced Photon Source (APS), Argonne National Laboratory. HPCAT is supported by CIW, CDAC, UNLV and LLNL through funding from DOE-NNSA, DOE-BES and NSF. APS is supported by DOE-BES, under DE-AC02-06CH11357. The UNLV HPSEC was supported by the U.S. DOE, National Nuclear Security Administration, under DE-FC52-06NA26274.

**Table I:** Unit-cell parameters and atomic coordinates for ScAlMgO$_4$ obtained from powder diffraction at RT and ambient pressure.

| \multicolumn{5}{c}{$a$ = 3.245(1) Å and $c$ = 25.160(9) Å; V = 229.4(2) Å$^3$; Z = 3} |
|---|---|---|---|---|
| Atom | Site | x | y | z |
| Sc | 3a | 0 | 0 | 0 |
| Al/Mg | 6c | 0 | 0 | 0.217(1) |
| O$_1$ | 6c | 0 | 0 | 0.128(1) |
| O$_2$ | 6c | 0 | 0 | 0.293(1) |

**Table II:** Unit-cell parameters and atomic coordinates for ScAlMgO$_4$ obtained from single-crystal diffraction at RT and ambient pressure.

| \multicolumn{6}{c}{$a$ = 3.25385(8) Å and $c$ = 25.2318(4) Å; V = 231.35(1) Å$^3$; Z = 3} |
|---|---|---|---|---|---|
| Atom | Site | x | y | z | U$_{iso}$ |
| Sc | 3a | 0 | 0 | 0 | 0.0106 |
| Al/Mg | 6c | 0 | 0 | 0.216453 | 0.0085 |
| O$_1$ | 6c | 0 | 0 | 0.127713 | 0.0232 |
| O$_2$ | 6c | 0 | 0 | 0.293008 | 0.0100 |

| Data Collection |
|---|
| Total reflections: 1178 |
| -4 ≤ h ≤ 4, -4 ≤ k ≤ 4, -32 ≤ l ≤ 32 |
| Max. 2θ = 57.42° |
| Unique reflections: 103 |
| R1 = 0.0189 |
| wR2 = 0.0548 |

**Table III:** Unit-cell parameters and atomic coordinates for ScAlMgO$_4$ obtained from single-crystal diffraction at 100 K and ambient pressure.

| $a = 3.249(1)$ Å and $c = 25.187(4)$ Å; $V = 230.25(1)$ Å$^3$; $Z = 3$ | | | | | |
|---|---|---|---|---|---|
| Atom | Site | x | y | z | U$_{iso}$ |
| Sc | 3a | 0 | 0 | 0 | 0.0087 |
| Al/Mg | 6c | 0 | 0 | 0.216371 | 0.0071 |
| O$_1$ | 6c | 0 | 0 | 0.127796 | 0.0216 |
| O$_2$ | 6c | 0 | 0 | 0.292961 | 0.0088 |

Data Collection

Total reflections: 1178

$-4 \leq h \leq 4$, $-4 \leq k \leq 4$, $-34 \leq l \leq 33$

Max. $2\theta = 59.50°$

Unique reflections: 111

R1 = 0.0196

wR2 = 0.0566

**Table IV:** Unit-cell parameters and atomic coordinates for high-pressure ScAlMgO$_4$ obtained from powder diffraction at RT and 28 GPa.

| $a = 16.07(5)$ Å, $b = 3.15(1)$ Å, $c = 8.02(2)$ Å, and $\beta = 160.8(2)°$; $V = 229.4(2)$ Å$^3$; $Z = 2$ | | | | |
|---|---|---|---|---|
| Atom | Site | x | y | z |
| Sc | 2e | 0 | 0 | 0 |
| Al/Mg | 4i | 0.782(3) | 0 | 0.217(1) |
| O$_1$ | 4i | 0.872(4) | 0 | 0.128(1) |
| O$_2$ | 4l | 0.707(3) | 0 | 0.293(2) |

**Figure Captions**

**Figure 1:** Schematic view of the layered structure of $ScAlMgO_4$.

**Figure 2:** Selection of x-ray diffraction patterns collected up to 24 GPa ($\lambda$ = 0.41514 Å). Background was subtracted. Ne peaks are labeled and pressures indicated. Peaks of $ScAlMgO_4$ mentioned in the discussion are indexed. The dotted lines illustrate the different pressure evolution of (006) and (104) reflections and the splitting of (116) and (0114) reflections.

**Figure 3:** Unit-cell parameters, volume, and axial ratio as a function of pressure. Different symbols correspond to different experiments. The solid lines are cubic fits to the data with the exemption of the volume plot where we plotted the fitted EOS.

**Figure 4:** Pressure dependence of the bond distances. The incompressible intralayer Al/Mg-O bond is shown with empty circles.

**Figure 5:** X-ray diffraction patterns collected at different pressures for the low- and high-pressure phases ($\lambda$ = 0.40753 Å). Background was subtracted. The bottom and top patterns correspond to the low-pressure phase. The other two patterns to the high-pressure phase. The ticks indicate the positions of Ne and $ScAlMgO_4$. (r) denotes data collected upon pressure release.

**Figure 1**

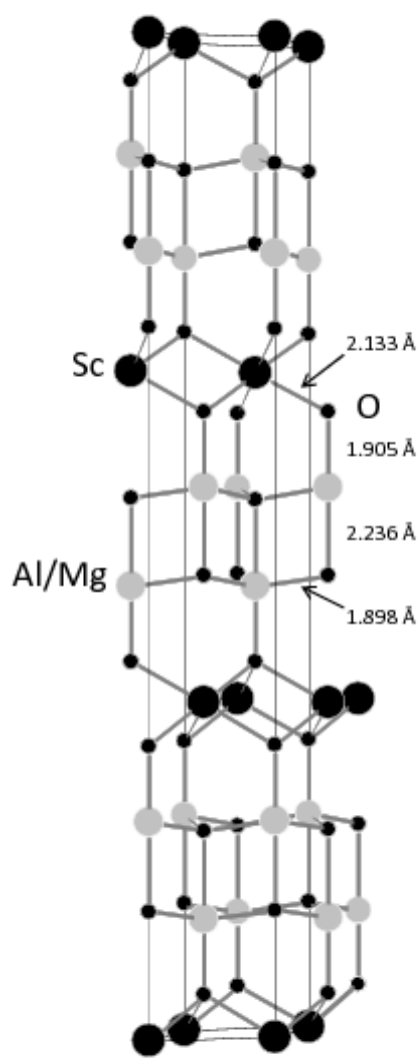

**Figure 2**

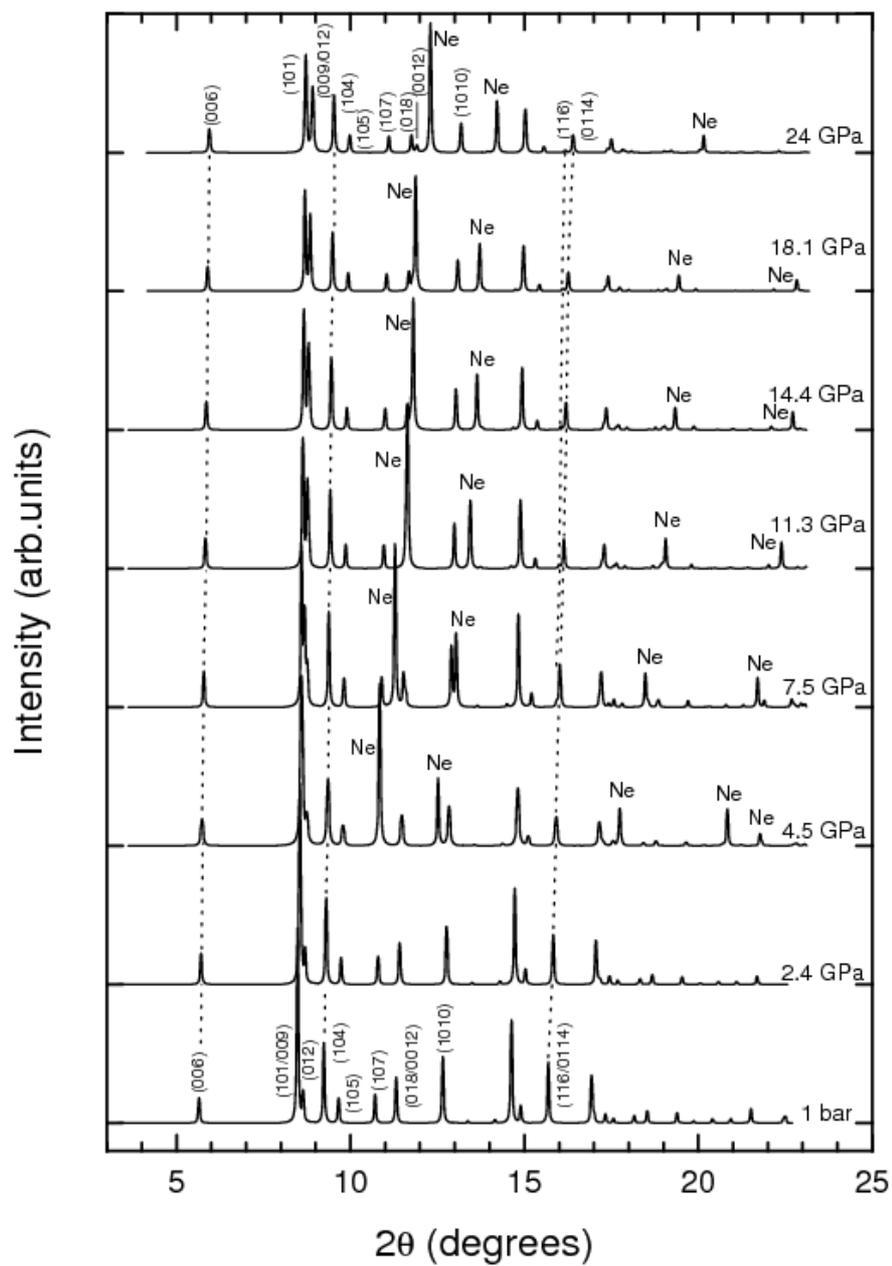

**Figure 3**

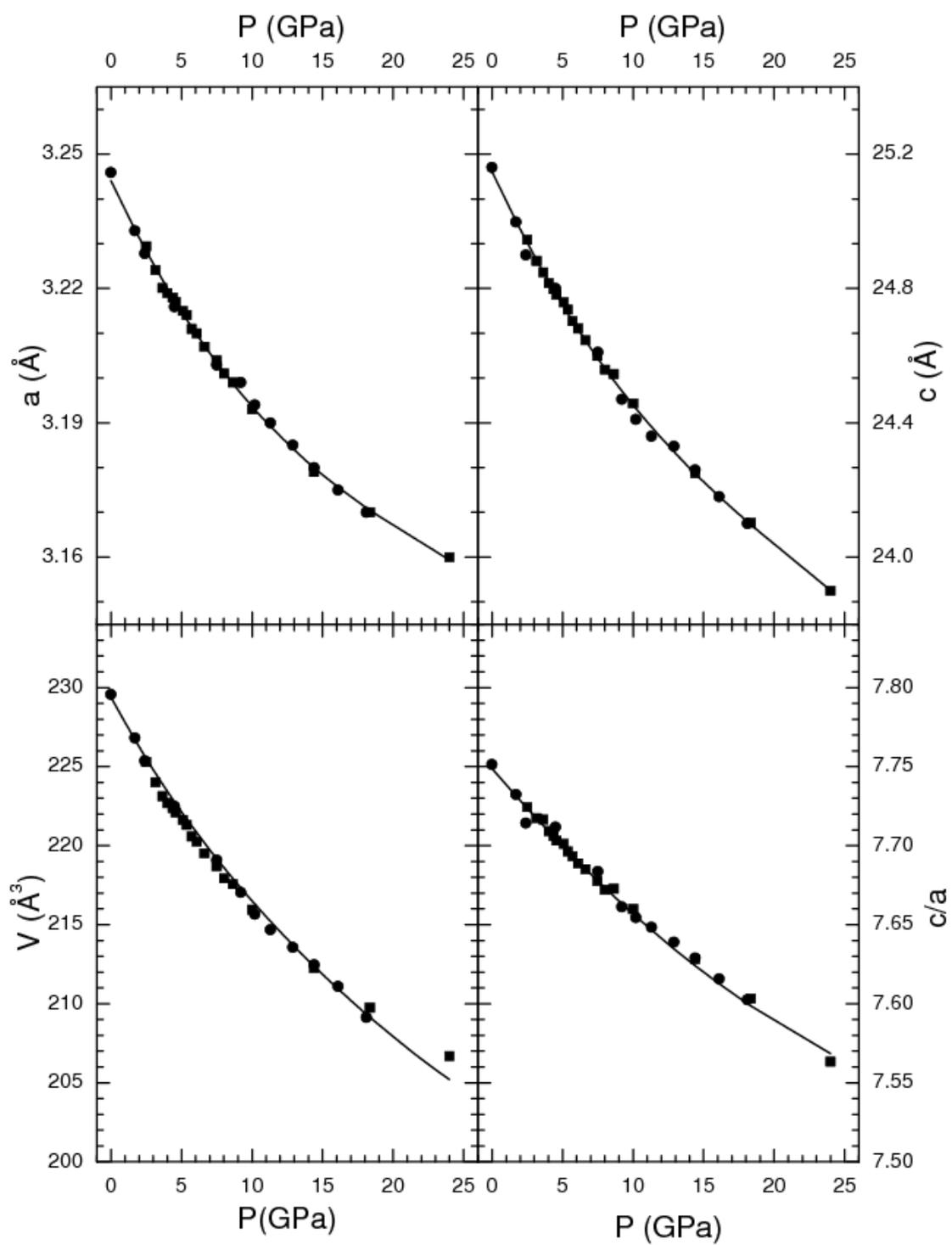

**Figure 4**

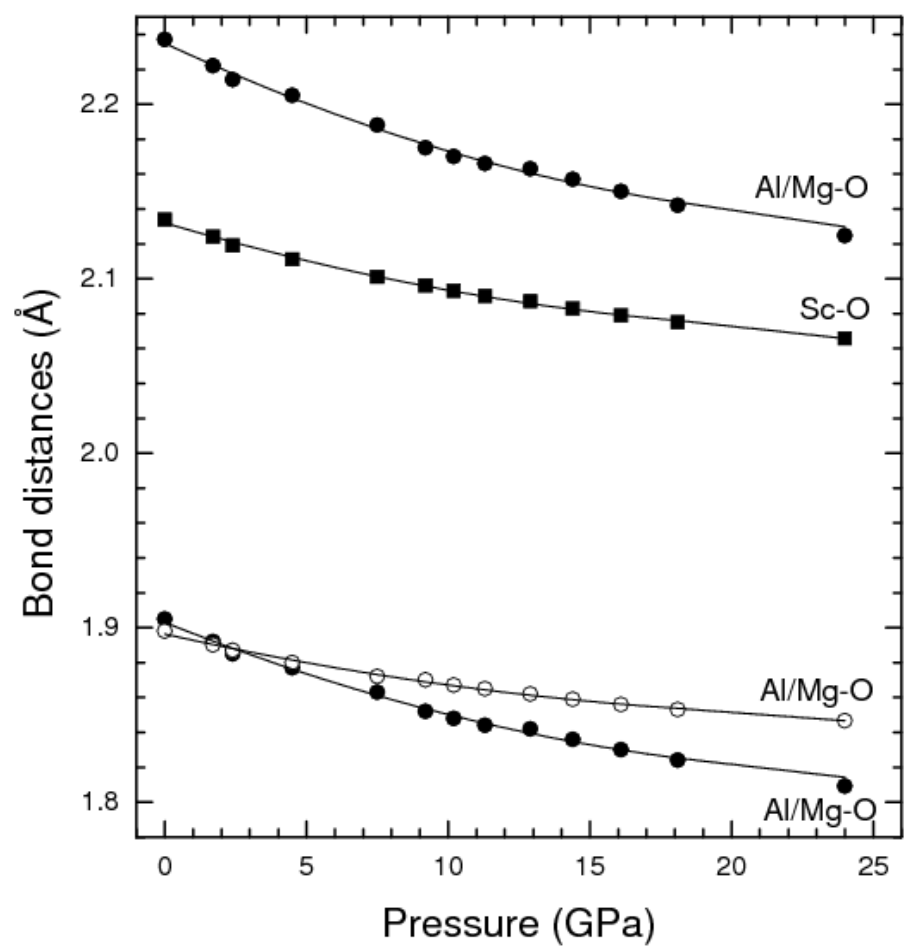

**Figure 5**

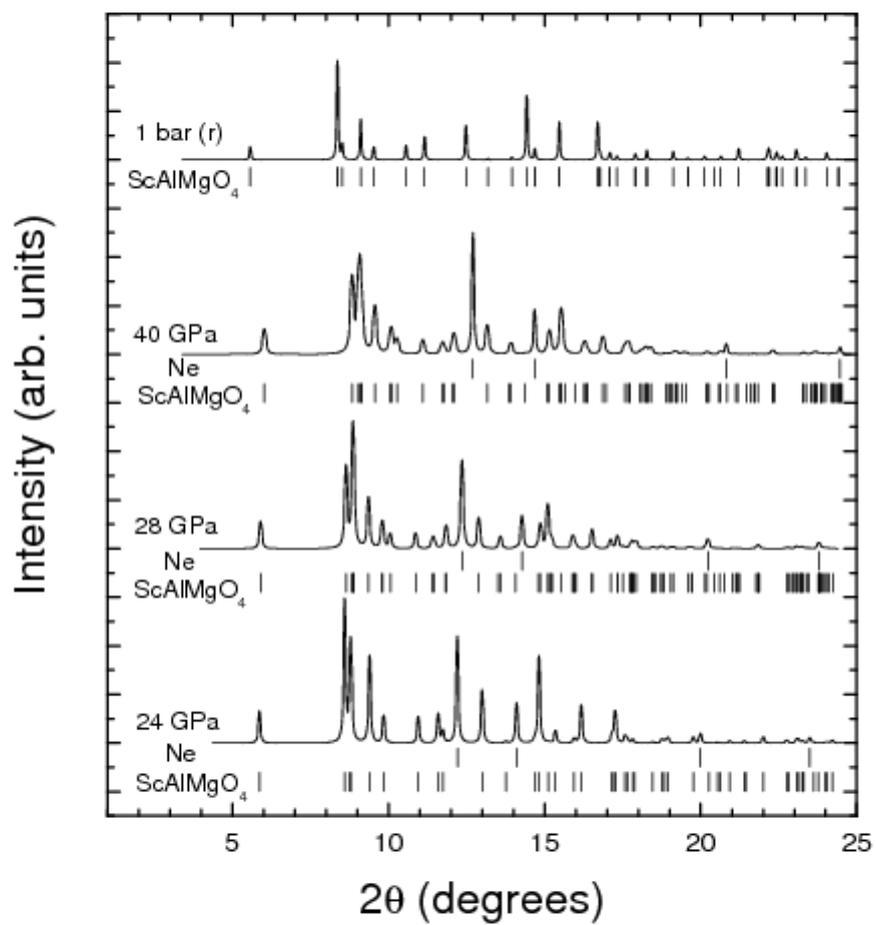